\documentclass[runningheads]{svjour2}
\smartqed  
\usepackage{graphicx}
\usepackage[numbers]{natbib}
\usepackage{mathptmx}
\usepackage{biograph}        
\newcommand{\vd}{v_{\rm d}}            
\newcommand{\DR}{D^{\rm R}}            
\newcommand{\zf}{z_{\rm f}}            
\newcommand{\tf}{{t_{\rm f}}}          
\newcommand{\ap}{\alpha_{\rm p}}       

\DeclareMathAlphabet{\bs}{OT1}{ptm}{b}{it}

\journalname{Journal of Mathematical Biology}
\begin{document}

\title{Persistence of direction increases the drift velocity of run and
tumble chemotaxis
\thanks{This work was supported by an Oliver Gatty Studentship from
the University of Cambridge.}
}

\titlerunning{Persistence enhances chemotaxis}        

\author{J. T. Locsei}


\institute{J. T. Locsei \at
              Department of Applied Mathematics and Theoretical Physics,
              University of Cambridge, Cambridge CB3 0WA, U.K. \\
              \email{j.t.locsei@damtp.cam.ac.uk}           
}

\date{Received: date / Revised: date}

\maketitle

\begin{abstract}
\emph{Escherichia coli} is a motile bacterium that moves up a
chemoattractant gradient by performing a biased random walk composed of
alternating runs and tumbles. Previous models of run and tumble
chemotaxis neglect one or more features of the motion, namely (i) a
cell cannot directly detect a chemoattractant gradient but rather makes
temporal comparisons of chemoattractant concentration, (ii) rather than
being entirely random, tumbles exhibit persistence of direction,
meaning that the new direction after a tumble is more likely to be in
the forward hemisphere, and (iii) rotational Brownian motion makes it
impossible for an \emph{E. coli} cell to swim in a straight line during
a run. This paper presents an analytic calculation of the chemotactic
drift velocity taking account of (i), (ii) and (iii), for weak
chemotaxis. The analytic results are verified by Monte Carlo
simulation. The results reveal a synergy between temporal comparisons
and persistence that enhances the drift velocity, while rotational
Brownian motion reduces the drift velocity.

\end{abstract}

\section{Introduction}
\label{sec:intro}

As described by \citet{Berg.1983}, \emph{Escherichia coli} is a common
intestinal bacterium with a body $\approx 1\;\mu{\rm m}$ in diameter
and $\approx 2\;\mu{\rm m}$ long. A typical \emph{E. coli} cell
possesses approximately 6 helical flagella, each $\approx 20{\rm\;nm}$
thick and $\approx 10 \; \mu {\rm m}$ long, which emerge from random
points on the cell membrane. Each flagellum is powered by a reversible
rotary motor. When the flagella spin counter-clockwise, they form a
synchronous bundle and propel the cell in an approximately straight
path called a `run', with a swimming velocity of $\approx 20 \; \mu
{\rm m\,s^{-1}}$ and a corresponding Reynolds number of $\textit{Re}
\approx 10^{-5}$. The run is not entirely straight, as the cell is
subject to rotational and translational Brownian motion due to thermal
collisions with the molecules in the surrounding fluid. Translational
Brownian motion is negligible compared to the cell's swimming motion,
but rotational Brownian motion is significant and in water at room
temperature it causes a root-mean-square angular deviation in radians
of about $0.5 \,t^{1/2}$, where $t$ is in seconds. When one or more
flagella spin clockwise, the bundle comes apart and the cell moves in
an erratic path with little net displacement called a `tumble', which
reorients the cell to face a random new direction. Tumbles exhibit
persistence of direction (hereafter simply referred to as
`persistence'), meaning that the direction faced by the cell after a
tumble is not isotropically random. Rather, the mean angle between the
new and the previous direction is $62^{\rm o}$ \citep{Berg.1972}, so
the new direction after a tumble is more likely to be in the forward
hemisphere.

An \emph{E. coli} cell performs chemotaxis by executing a biased random
walk composed of alternating runs and tumbles. The run durations are
well described by a Poisson interval distribution (\emph{i.e.}
exponential distribution) with a rate constant (`tumble rate') of
$\lambda \approx 1 \; {\rm s^{-1}}$, and a corresponding mean run
duration of $\lambda^{-1} \approx 1 \; {\rm s}$ \citep{Berg.1972}.
Tumble durations also follow a Poisson interval distribution, but are
much shorter than runs, with a mean duration of $\approx 0.1 \; {\rm
s}$. An \emph{E. coli} cell is too small to detect spatial differences
in the concentration of a chemoattractant on the scale of a cell
length. Instead, the cell performs temporal comparisons. As the cell
swims about it continually measures the concentration of
chemoattractants in its environment (\emph{e.g.} serine, aspartate) and
compares the chemoattractant concentration over the past second with
the concentration over the previous three seconds. The cell modulates
the tumble rate in response to a concentration difference, so that runs
up the chemoattractant gradient are extended. The direction-dependent
tumble rate causes the cell to drift toward regions of high
chemoattractant concentration.

One measure of the effectiveness of chemotaxis is the drift velocity,
defined as the mean velocity at which a cell moves up a chemoattractant
gradient. No previously published model of chemotaxis allows the
calculation of the chemotactic drift velocity while simultaneously
accounting for temporal comparisons, persistence, and rotational
Brownian motion. Existing literature on the topic includes the
following. \citet{Patlak.1953} presented an analysis of random walks
with persistence and bias, but his framework does not accommodate
temporal comparisons. \citet{Brown.1974} performed Monte Carlo
simulations for a model in which a cell possesses an exponentially
decaying memory; they included persistence but not rotational Brownian
motion. \citet{Schnitzer.1993} presented a theory of random walks with
persistence and rotational Brownian motion, but without temporal
comparisons. \citet{deGennes.2004} presented an analytic method for
calculating the chemotactic drift velocity taking account of temporal
comparisons, but neglected Brownian motion and persistence. Recently,
\citet{Erban.2005} achieved the noteworthy task of deriving a complete
advection-diffusion equation for a suspension of chemotactic organisms
taking account of both temporal comparisons and persistence.

In this paper we extend the model of \citet{deGennes.2004} to include
Brownian motion and persistence, and we report a new effect:
persistence can markedly enhance the drift velocity.

\section{The model}
\label{sec:model}

Consider a cell performing run and tumble chemotaxis, swimming in an
unbounded, stationary fluid which contains a uniform chemoattractant
concentration gradient \mbox{$\nabla c \parallel \hat{\bs{z}}$}, where
$\hat{\bs{z}}$ denotes the unit vector in the $z$ direction. During a
run, the cells swims at constant speed $v_{\rm s}$, and the probability
that the cell tumbles in the next time interval $dt$ is $\lambda(t)
dt$, where $\lambda$ is the `tumble rate'. Like \citet{deGennes.2004},
we assume that for weak concentration gradients the tumble rate depends
linearly on chemoattractant concentration history, so that the tumble
rate is given by:
\begin{equation}
\label{eq:lambda}
   \lambda(t)=\lambda_0
      \left[
         1 - \Delta(t)
      \right],
\end{equation}
where $\lambda_0 = 1 {\rm s}^{-1}$ is the baseline tumble rate, and the
fractional change in tumble rate is given by
\begin{equation}
\label{eq:Delta}
  \Delta(t)=\int_{-\infty}^t c(t') R(t-t') dt',
\end{equation}
where $c(t')$ is the chemoattractant concentration experienced by the
cell at time $t'$, and $R$ is the cell's `response function'. Note that
$R$ may be thought of as the impulse response of the tumble rate, since
it describes the way that $\lambda$ changes when the cell is subject to
a Dirac delta impulse of chemoattractant concentration. Equation
(\ref{eq:lambda}) also appears in the papers of \citet{Schnitzer.1993}
and \citet{Clark.2005}, and it is motivated by the experimental results
of \citet{Block.1982} and \citet{Segall.1982}. Our analysis will be
restricted to `weak chemotaxis', meaning small fractional changes in
the tumble rate, \emph{i.e.} $|\Delta(t)| \ll 1$. Physically, weak
chemotaxis corresponds to a shallow chemoattractant gradient.

Rotational Brownian motion causes the cell to gradually veer off
course, so that in between tumbles the probability density function $f$
of the swimming direction $\bs{e}$ evolves according to the
Fokker-Planck equation
\begin{equation}
\label{eq:fpe}
   \frac{\partial f}{\partial t} = \DR \nabla_{\bs{e}}^2 f,
\end{equation}
where $\DR$ is the rotational diffusion coefficient and
$\nabla_{\bs{e}}^2$ is the Laplacian in direction space.
\citet{Berg.1983} estimated $\DR \approx 0.062\;{\rm s}^{-1}$ for an
\emph{E. coli} cell swimming in water at room temperature.

When a cell tumbles, its choice of new direction is governed by a
probability distribution which is axisymmetric about the initial
direction. We allow for the tumbles to exhibit directional persistence,
so that the expected scalar product of the swimming directions
$\bs{e}(0^-)$ and $\bs{e}(0^+)$ immediately before and after a tumble
at time $t=0$ is given by
\begin{equation}
\label{eq:alphap}
   E[ \bs{e}(0^-) \cdot \bs{e}(0^+) ] = \ap,
\end{equation}
where $\ap$ is the `persistence parameter'. Experimentally, $\ap
\approx 0.33$ \citep{Berg.1983}.

The drift velocity, $\vd$, can be defined in terms of the expected
motion in a single run. Let $z_{\rm f}$ be the $z$ location of a cell
at the end of a run, relative to its position at the beginning of a
run, and let $t_{\rm f}$ be the duration of a run. We treat the
duration of tumbles as negligible, so
\begin{equation}
\label{eq:vd}
   \vd = E[ \zf ] / E[ \tf ]
\end{equation}
where $E$ denotes an expectation value. If the non-zero duration of
tumbles is taken into account, then the only effect is to reduce
$v_{\rm d}$ by a small factor. In taking expectation values, we assume
that the cell has already been swimming in the fluid for a sufficiently
long time that the probability density function of $z_{\rm f}$ is
time-independent. While our model assumes an unbounded domain, we note
that for a cell swimming in a bounded domain of length $\gg v_{\rm s} /
\lambda_0$, $\vd$ provides a measure of the transient average velocity
up the chemoattractant gradient before the cell encounters the
boundaries.

\section{Calculation of drift velocity for general response function}
\label{sec:driftcalc}

Calculating $\vd$ is non-trivial because of the interdependence of the
tumble rate and the path taken by the cell. The tumble rate at any time
depends in principle on the entire path history of the cell through
(\ref{eq:Delta}), while the path of the cell depends in turn on the
tumble rate. As noted above, in order to make the analysis tractable,
our analysis will be restricted to weak chemotaxis, \emph{i.e.}
\mbox{$0\leq|\Delta(t)| \ll 1$}. In this case, one has $E[\zf] = v_{\rm
s} \lambda_0 O(\Delta)$ and $E[\tf] = [1+O(\Delta)]/\lambda_0$, so
\begin{equation} \label{eq:vd1}
   \vd = \lambda_0 E[\zf] + v_{\rm s} O(\Delta^2).
\end{equation}
We shall neglect terms that are $O(\Delta^2)$.

Consider a run commencing at time $t = 0$ at location $z = 0$. During
the run, the cell swims in a random walk governed by rotational
Brownian motion until the run terminates with a tumble at time $\tf$.
Note that the expected stopping location of a terminated random walk
with a stopping rate $\lambda_{\rm stop}$ is the same as the expected
first event location on an unterminated walk with an event rate
$\lambda_{\rm event} = \lambda_{\rm stop}$. Thus, in calculating
$E[\zf]$, it is permissible to treat tumbles for $t>0$ as events that
have no effect on the cell's motion, and treat the tumble at $t = \tf$,
$z=\zf$ as a first event (the first tumble in $t>0$). The utility of
this treatment is that we may conceptually break the expectation $E$ in
$E[\zf]$ into two consecutive operations. First, assuming a given path
$z(t):-\infty<t<\infty$ taken by the cell, one calculates the
conditional expectation of $\zf$ for that path. Second, one takes the
expectation over all such paths to obtain $E[\zf]$, with the
understanding that in the $t\leq 0$ section of a path the cell is
subject to reorientations due to both Brownian motion and tumbles,
whereas in the $t>0$ section of a path the cell is subject to
reorientation due to Brownian motion alone. Writing out the two
expectations in symbolic notation,
\begin{equation}
\label{eq:zf0}
   E[ \zf ] = E_{\rm paths}\left[\int_0^{\infty} dt\,
   z_{\rm path}(t) p_{\rm path}(t) \right],
\end{equation}
where the $E_{\rm paths}$ denotes an expectation over paths, $z_{\rm
path}(t)$ denotes the position of the cell at time $t$ on a particular
path, and $p_{\rm path}(t)$ is the probability density function for the
tumble time $\tf$ on a particular path. Since tumbles for $t>0$ are
treated as having no effect on cell motion, paths are independent of
$\tf$ and one is free to take the path expectation inside the integral
over tumble times. Dropping the `path' subscript for brevity, one then
has
\begin{equation}
\label{eq:zf1}
   E[ \zf ] = \int_0^{\infty} dt\, E[z(t) p(t)],
\end{equation}

The probability density function p(t) for the tumble time is given by
\begin{equation}
\label{eq:ptf}
   p(t) = \lambda(t)
   \exp\left[-\int_0^{t} \lambda(t')\,dt'\right],
\end{equation}
where $\lambda(t)$ is the path-dependent tumble rate at time $t$.
Substituting (\ref{eq:ptf}) into (\ref{eq:zf1}) and (\ref{eq:vd1}) and
integrating by parts yields
\begin{equation}
\label{eq:vd2}
   \vd = \lambda_0
    \int_0^\infty dt \,E\left[ w(t)
   \exp\left[-\int_0^{t} \lambda(t')\,dt'\right]
   \right],
\end{equation}
where $w(t)=d\,z(t)/dt$. Writing $\lambda(t) = \lambda_0[1-\Delta(t)]$
in (\ref{eq:vd2}), expanding the exponential in powers of $\Delta$ and
keeping only the linear term, one obtains
\begin{equation}
\label{eq:vd3}
   \vd = \lambda_0
    \int_0^\infty dt \,E\left[ w(t)
    {\rm e}^{-\lambda_0 t}
    \left[1+\lambda_0 \int_0^{t} \Delta(t')\,dt'\right]
   \right].
\end{equation}

We shall calculate the drift velocity for the case where $R$ is given
by a Dirac delta function and later generalise to an arbitrary response
function. Consider the response function
\begin{equation}
   R(t)=A\,\delta(t-T).
\end{equation}
With this $R$, the fractional change in tumble rate is simply
\begin{eqnarray}
\label{eq:delta2} \nonumber
   \Delta(t) &=& A \,c(t-T)] \\
   &=& A\, c_0 + A |\nabla c|z(t-T)],
\end{eqnarray}
where $c(t)= c_0+|\nabla c|z(t)$ is the chemoattractant concentration
seen by the cell at time $t$. We shall discard the $c_0$ term, since
eventually we shall require the response function $R$ to have zero
mean, so that an additive constant to $c$ in (\ref{eq:Delta}) has no
effect on $\Delta$. Substituting (\ref{eq:delta2}) into (\ref{eq:vd3})
and discarding terms containing $c_0$, one finds

\begin{equation} \label{eq:vd4}
   \vd =\lambda_0 \int_0^\infty dt \, {\rm e}^{-\lambda_0 t} E[w(t)]
   + \lambda_0^2 A |\nabla c| \int_0^\infty dt \,
   {\rm e}^{-\lambda_0 t} \int_0^{t}dt'\,E[w(t)z(t'-T)].
\end{equation}

We may write
\begin{equation} \label{eq:Ewtstart}
   E[w(t)] = v_{\rm s} \hat{\bs{z}} \cdot E[\bs{e}(t)].
\end{equation}
The probability density function for the angular deviation caused by
rotational Brownian motion is symmetric about initial swimming
direction, so we may write
\begin{equation} \label{eq:Ewtfderivation}
   E[\bs{e}(t)]=E[\bs{e}(t)\cdot\bs{e}(0^+)]E[\bs{e}(0^+)]
\end{equation}
where $\bs{e}(0^+)$ is the cell's swimming direction at the beginning
of the run. One can straightforwardly solve the Fokker-Planck equation
(\ref{eq:fpe}) to find that the direction correlation function in the
absence of tumbles is
\begin{equation}
\label{eq:dircort1t2}
   E[ \bs{e}(t_1) \cdot \bs{e}(t_2) ] =
   {\rm e}^{-2\DR|t_2-t_1|}
\end{equation}
($t_1,t_2>0$), and substitution of (\ref{eq:dircort1t2}) into
(\ref{eq:Ewtfderivation}) and (\ref{eq:Ewtstart}) gives
\begin{equation} \label{eq:Ewtf}
   E[w(t)] = {\rm e}^{-2\DR t} E[w(0^+)],
\end{equation}
where $w(0^+)$ is the $z$ component of the cell's velocity at the
beginning of the run. Substituting (\ref{eq:Ewtf}) into (\ref{eq:vd4})
and swapping the order of integration over $t$ and $t'$ yields
\begin{eqnarray} \label{eq:vd5}
\nonumber
   \vd &=& E[w(0^+)] \lambda_0 /(\lambda_0+2\DR)\\
   && + \lambda_0^2 A |\nabla c| \int_0^\infty dt'
   \int_{t'}^\infty dt\, {\rm e}^{-\lambda_0 t} E[w(t) z(t'-T)]
\end{eqnarray}

To proceed further, we must find expressions for $E[ w(0^+) ]$ and
$E[w(t_{\rm b}) z(t_{\rm a})]$, for $-\infty < t_{\rm a} \leq t_{\rm
b}$ and $0 \leq t_{\rm b} < \infty$. Since we are neglecting terms
$O(\Delta^2)$ in $\vd$, and the $A |\nabla c|$ factor in (\ref{eq:vd5})
comes from one power of $\Delta$, we can neglect chemotaxis altogether
in calculating $E[w(t_{\rm b}) z(t_{\rm a})]$. Neglecting chemotaxis,
position and velocity are governed by isotropic distributions, so
\begin{eqnarray}
\label{eq:g2} \nonumber
   E[w(t_{\rm b}) z(t_{\rm a})] &=& \int_0^{t_{\rm a}} dt\,E[w(t_{\rm b}) w(t)]
   \\
   &=& \frac{1}{3} v_{\rm s}^2
   \int_0^{t_{\rm a}} dt\, E[\bs{e}(t_{\rm b}) \cdot \bs{e}(t)].
\end{eqnarray}
For $0<t_{\rm a}<t_{\rm b}$, we can use equation (\ref{eq:dircort1t2})
for the direction correlation function to find that
\begin{eqnarray}
\label{eq:gtpos} \nonumber
   E[w(t_{\rm b}) z(t_{\rm a})] &=& \frac{1}{3}  v_{\rm s}^2 \int_0^{t_{\rm a}} dt {\rm e}^{2\DR(t-t_{\rm b})} \\
   &=& \frac{v_{\rm s}^2}{6\DR}\,{\rm e}^{-2\DR t_{\rm b}}
   \left({\rm e}^{2\DR t_{\rm a}}-1\right),
   \; 0 <t_{\rm a} \leq t_{\rm b}.
\end{eqnarray}
For $t_{\rm a}<0$ [and hence $t<0$ in (\ref{eq:g2})], we expand the
direction correlation function into a product of direction correlations
between different times, and then simplify using (\ref{eq:alphap}) and
(\ref{eq:dircort1t2}):
\begin{eqnarray} \label{eq:Etftp}
\nonumber
   E[\bs{e}(t_{\rm b}) \cdot \bs{e}(t)] &=&
      E[\bs{e}(t_{\rm b}) \cdot \bs{e}(0^+)]\,
      E[\bs{e}(0^+) \cdot \bs{e}(0^-)] \,
      E[\bs{e}(0^-) \cdot \bs{e}(t)]\\
   &=& \ap {\rm e}^{-2\DR t_{\rm b}}\,E[\bs{e}(0^-) \cdot \bs{e}(t)].
\end{eqnarray}
To obtain an expression for $E[\bs{e}(0^-) \cdot \bs{e}(t)]$ we note
that for \mbox{$t<0$} the cell undergoes reorientations due to both
rotational Brownian motion and tumbles. \citet{Lovely.1975} showed that
the direction correlation function for isotropic run and tumble random
walks without rotational Brownian motion is
\begin{equation} \label{eq:dircor2}
   E[ \bs{e}(t_1) \cdot \bs{e}(t_2) ]= {\rm e}^{-|t_1-t_2|/\tau},
\end{equation}
where
\begin{equation}
\label{eq:tau}
   \tau = \frac{1}{\lambda_0(1-\ap)}
\end{equation}
is the direction correlation time. \citet{Lovely.1975} also showed that
rotational Brownian motion can be regarded as a Poisson tumbling
process, and that if two Poisson tumbling processes have individual
direction correlation times $\tau_1$ and $\tau_2$, then the direction
correlation for the combined process is still of the form
(\ref{eq:dircor2}), but with correlation time $\tau_c$ given by
\begin{equation}
\label{eq:tau_c}
   \frac{1}{\tau_c} = \frac{1}{\tau_1} + \frac{1}{\tau_2}.
\end{equation}
One may think of $\tau_c$ as the characteristic time it takes the cell
to perform a `U-turn'. We apply (\ref{eq:tau}) and (\ref{eq:tau_c}) to
our problem to find that
\begin{equation} \label{eq:tau_c2}
   \tau_c = 1/[\lambda_0(1-\ap) + 2\DR]
\end{equation}
and
\begin{equation} \label{eq:dircor3}
   E[ \bs{e}(0^-) \cdot \bs{e}(t) ]
   = {\rm e}^{[\lambda_0(1-\ap) + 2\DR]t}.
\end{equation}
Substitution of (\ref{eq:dircor3}) and (\ref{eq:Etftp}) into
(\ref{eq:g2}) yields
\begin{equation}
\label{eq:gtneg}
   E[w(t_{\rm b}) z(t_{\rm a})] =
   \frac{
      \ap v_{\rm s}^2 {\rm e}^{-2\DR t_{\rm b}} ( {\rm e}^{[\lambda_0(1-\ap) + 2\DR]t_{\rm a}}-1 )
   }{3[\lambda_0(1-\ap) + 2\DR]}, \; t_{\rm a} \leq 0 \leq t_{\rm b}.
\end{equation}

Having found $E[w(t_{\rm b}) z(t_{\rm a})]$, we turn our attention to
$E[ w(0^+) ]$. In the absence of rotational Brownian motion, $E[ w(0^+)
] = 0$, since although the cell alters run durations in response to the
chemoattractant, run directions remain random and isotropic. In the
presence of Brownian motion, however, this is no longer true. During a
run, rotational Brownian motion causes the cell's swimming direction to
drift. For instance, modulation of the tumble rate might make it more
likely that the cell tumbles when heading down the chemoattractant
gradient, so there is down-gradient bias in the expected swimming
direction at the end of a run. If $\ap>0$ the cell is consequently more
likely to commence each run heading down-gradient.

From the definition (\ref{eq:alphap}) of the persistence parameter, we
may write
\begin{equation}
   E[w(0^+)] = \ap E[w(0^-)].
\end{equation}
Furthermore, if we assume that the cell has been swimming in the
chemoattractant gradient for a time $t_{\rm swim} \gg \tau_{\rm c}$
[where $\tau_c$ is given by (\ref{eq:tau_c2})], then the expected
swimming velocity just before a tumble does not change from one tumble
to the next, so $E[w(\tf^-)] = E[w(0^-)]$ and hence
\begin{equation} \label{eq:Ew0+}
   E[w(0^+)] = \ap E[w(\tf^-)].
\end{equation}
As in the derivation of (\ref{eq:zf1}), we may break the expectation on
the right hand side of (\ref{eq:Ew0+}) into an integral over tumble
times and an expectation over paths:
\begin{equation} \label{eq:Ewtf-}
   E[w(\tf^-)] = \int_0^t dt\, E[w(t)p(t)]
\end{equation}
Using (\ref{eq:ptf}) and (\ref{eq:lambda}) to expand $p(t)$ in powers
of $\Delta$ and keeping only the linear term, one obtains
\begin{equation} \label{eq:pt2}
   p(t)= \lambda_0{\rm e}^{-\lambda_0 t}
   \left[ 1-A |\nabla c| z(t-T)
   +\lambda_0 A |\nabla c| \int_0^{t}z(t'-T)dt'
   \right].
\end{equation}
Substituting (\ref{eq:pt2}) into (\ref{eq:Ewtf-}) and re-arranging
yields
\begin{equation} \label{eq:Ewtf-2}
   E[w(\tf^-)] =
   \lambda_0 \int_0^\infty dt \,{\rm e}^{-\lambda_0 t}
   E[w(t)] - \lambda_0 A |\nabla c| (I_1-I_2),
\end{equation}
where
\begin{equation}
   I_1 = \int_0^\infty dt \,{\rm e}^{-\lambda_0 t} E[w(t),z(t-T)],
\end{equation}
and
\begin{equation}
   I_2 =\lambda_0 \int_0^\infty dt'
   \int_{t'}^{\infty} dt \, {\rm e}^{-\lambda_0 t} E[w(t),z(t'-T)].
\end{equation}
Using (\ref{eq:Ewtf}), (\ref{eq:Ewtf-2}) simplifies to
\begin{equation}
\label{eq:Ewtf-3}
   E[w(\tf^-)] = E[w(0^+)] \lambda_0 /(\lambda_0+2\DR)
   - \lambda_0 A |\nabla c| (I_1-I_2).
\end{equation}
Eliminating $E[w(\tf^-)]$ from (\ref{eq:Ew0+}) and (\ref{eq:Ewtf-3})
then yields
\begin{equation}
\label{eq:wplus}
   E[w(0^+)] =
   \frac{
      \lambda_0 A |\nabla c| \ap (2\DR+\lambda_0)(I_2-I_1)
   }{
      \lambda_0(1-\ap)+2\DR
   }.
\end{equation}
Using the definitions of $I_1$ and $I_2$ and equations (\ref{eq:gtneg})
and (\ref{eq:gtpos}) for $E[w(t_{\rm b}),z(t_{\rm a})]$, one finally
obtains the expression for $E[w(0^+)]$:
\begin{equation}
\label{eq:Ew0+2}
   E[ w(0^+)] = -
   \frac{
      2 A |\nabla c| \ap \DR v_{\rm s}^2
      [(\lambda_0+2\DR){\rm e}^{-[\lambda_0(1-\ap)+2\DR]T}-\lambda_0 \ap]
   }{
      3(\lambda_0+2\DR)[\lambda_0(1-\ap)+2\DR]^2
   }.
\end{equation}

We now have all the necessary pieces to write down the expression for
$\vd$. Substituting (\ref{eq:Ew0+2}), (\ref{eq:gtneg})
and(\ref{eq:gtpos}) into (\ref{eq:vd5}), we obtain:
\begin{equation}
   \vd = A k(T),
\end{equation} where
\begin{equation}
\label{eq:kernel}
   k(T) =
   \frac{
      v_{\rm s}^2 |\nabla c| \lambda_0
      {\rm e}^{-(\lambda_0+2\DR)T}(1-\ap)
      [(\lambda_0+2\DR){\rm e}^{\lambda_0 \ap T}
      -\lambda_0 \ap {\rm e}^{(\lambda_0+2\DR)T}]
   }{
      3(\lambda_0+2\DR)[\lambda_0(1-\ap)+2\DR]^2
   }.
\end{equation}
The function $k$ is effectively the Greens function for an impulse
response function. For a general response function $R$, the drift
velocity is
\begin{equation}
\label{eq:vdsimpleintegral}
   \vd = \int_0^\infty dT \, R(T) k(T).
\end{equation}

Our calculations have been based on the assumption of a uniform
concentration gradient $\nabla c$. In fact, our results still apply to
the case of a non-uniform gradient, provided that the distance over
which $\nabla c$ varies is much larger than the distance covered by the
cell in a time $\tau_c$, \emph{i.e} provided that
\begin{equation}
   \left| \frac{\nabla c}{\nabla^2 c} \right|
   \gg v_{\rm s} \tau_c.
\end{equation}
Similarly, our calculations are based on the assumption that $c$ has no
time dependence in the Eulerian reference frame. However, our results
are still valid for time-dependent $c$, provided that the rate of
change of $c$ seen by a stationary cell is much slower than the rate of
change of $c$ seen by a moving cell \emph{i.e.} provided that
\begin{equation}
   \left|\frac{\partial c}{\partial t}\right| \ll
   v_{\rm s} |\nabla c|.
\end{equation}
Calculation of $\vd$ for time-dependent $c$ might be relevant, for
instance, to problems where the chemoattractant field is altered by
cells themselves through secretion or degradation.

\section{Consistency with known results}
\label{sec:equivalence}

\citet{deGennes.2004} derived an expression for the drift velocity in
the absence of rotational Brownian motion and persistence. Setting $\DR
= 0$ and $\ap = 0$ in (\ref{eq:kernel}) and (\ref{eq:vdsimpleintegral})
one finds
\begin{equation}
   \vd = \frac{v_{\rm s}^2 |\nabla c|}{3\lambda_0}
   \int_0^\infty dT \,R(T)\, {\rm e}^{-\lambda_0 T},
\end{equation}
consistent with equation 16 of \citep{deGennes.2004}.

\citet{Schnitzer.1993} derived expressions for the flux of cells in a
suspension of non-interacting chemotactic cells under the assumption
that each cell modifies its tumble rate according to its instantaneous
swimming direction relative to the chemoattractant gradient, so that
\begin{equation}
\label{eq:lambdaschnitz}
   \lambda = 1 - \epsilon \,\bs{e} \cdot \bs{e}_{\rm z}.
\end{equation}
We note that within our framework (\ref{eq:lambdaschnitz}) is
equivalent to setting
\begin{equation}
\label{eq:rschnitz}
   R(t) = \frac{\epsilon}{|\nabla c| v_{\rm s}}
   \frac{[\delta(t-T)-\delta(t-T - \Delta T)]}{\Delta T }
\end{equation}
and taking the simultaneous limits $\Delta T \rightarrow 0$ and $T
\rightarrow 0$. Substituting (\ref{eq:rschnitz}) into
(\ref{eq:vdsimpleintegral}) and taking $\Delta T \rightarrow 0$ and $T
\rightarrow 0$ yields
\begin{equation}
\label{eq:vdschnitz}
   \vd = \frac{\epsilon \lambda_0(1-\ap)}{3[2\DR+\lambda_0(1-\ap)]}.
\end{equation}
The flux $\bs{J}$ of cells in a uniform suspension is simply
proportional to the drift velocity. For cells swimming with unit
velocity,
\begin{equation}
\label{eq:Jschnitz}
   \bs{J} = \frac{\epsilon \lambda_0(1-\ap)}{3[2\DR+\lambda_0(1-\ap)]}
   \,\rho \bs{e}_{\rm z},
\end{equation}
where $\rho$ is the number density of cells. Equation
(\ref{eq:Jschnitz}) is consistent with equations 6.8 $(\ap>0,\DR=0)$
and 7.6 $(\DR>0,\ap=0)$ of \citep{Schnitzer.1993} in the case of a
uniform suspension. In passing, we note that \citet{Schnitzer.1993} did
not consider the case where both $\DR$ and $\ap$ are non-zero; in this
case (\ref{eq:Jschnitz}) shows that if $\DR > 0$ then $|\bs{J}|$ is a
decreasing function of $\ap$ and $|\bs{J}| \rightarrow 0$ as
$\ap\rightarrow 1$.

The work of \citet{Erban.2005} contains a derivation of an advection
diffusion equation for a suspension of chemotactic organisms, allowing
for temporal comparisons and persistence. Temporal comparisons are
modelled using internal states of the cell $y_1$ and $y_2$ that evolve
according to
\begin{eqnarray} \label{eq:y1y2}
\nonumber
   \dot{y_1} &=& \frac{f(c(t))-y_1-y_2}{t_{\rm e}} \\
   \dot{y_2} &=& \frac{f(c(t)) - y_2}{t_{\rm a}},
\end{eqnarray}
where $t_{\rm e}$ is an `excitation time' and $t_{\rm a} > t_{\rm e}$
is an `adaptation time', and $f(c)$ is the fraction of the cell's
receptors bound to chemoattractant molecules. The tumble rate is set as
$\lambda = \lambda_0 - b y_1$.

The \citet{Erban.2005} model for $\lambda(t)$ can be cast in the same
form as (\ref{eq:lambda}) and (\ref{eq:Delta}) by finding the general
solution to (\ref{eq:y1y2}):
\begin{equation}
   y_1(t) = \int_{-\infty}^t dt'\,f(c(t'))Y(t-t'),
\end{equation}
where
\begin{equation}
   Y(t) = \frac{1}{t_{\rm e}/t_{\rm a}}
   \left(
      \frac{{\rm e}^{-t/t_{\rm e}}}{t_{\rm e}} -
      \frac{{\rm e}^{-t/t_{\rm a}}}{t_{\rm a}}
   \right)
\end{equation}
is the Green's function for $y_1$. For small changes in concentration
around $c_0$, one then finds that the \citet{Erban.2005} model for
$\lambda(t)$ is equivalent to our model with a response function
\begin{equation} \label{eq:rErban}
   R(t) = \frac{b}{\lambda_0}f'(c_0) Y(t).
\end{equation}
Substitution of (\ref{eq:rErban}) into (\ref{eq:vdsimpleintegral}) then
yields a drift velocity of
\begin{equation}
   \vd =
   \frac{
      b\, t_{\rm a} v_{\rm s}^2 f'(c_0) |\nabla c|
   }{
      3 \lambda_0 [1+(1-\ap)\lambda_0 t_{\rm a}]
      [1+(1-\ap)\lambda_0 t_{\rm e}]
   },
\end{equation}
which is identical to the drift velocity that one can calculate from
equations 4.37 and 4.40 of \citep{Erban.2005}.

\section{Drift velocity for a `realistic' response function}

\subsection{Choice of response function}
\label{subsec:choosingR}

The validity of (\ref{eq:Delta}) has never been directly assessed, and
the chemotactic response function $R$ has never been directly measured.
However, \citet{Block.1982} and \citet{Segall.1986} monitored the
response of a single rotary motor on an \emph{E. coli} cell by
tethering the flagellum to a glass surface and delivering small
impulses of chemoattractants to the cell's immediate environment. By
repeating the experiment multiple times and with different cells, the
experimenters made measurements of the motor bias (the probability of
counter-clockwise rotation) as a function of time, (see for example
figure 1 of \citet{Segall.1986}). The impulse response of the motor
bias is double-lobed; the bias is raised above the baseline for the
first $\approx 1 \; {\rm s}$ after the delivery of the impulse, reduced
below the baseline for the following $\approx 3 \; {\rm s}$, and then
returns to baseline. The areas of the two lobes of the response are
equal. Furthermore, the experimenters found that the responses to other
time-series of stimuli (\emph{e.g.} ramp or sinusoid changes in
chemoattractant concentration) are consistent with the cell behaving as
a linear system, so that the response  to an arbitrary stimulus is well
described by the convolution integral of the stimulus with the impulse
response (\emph{c.f.} equation \ref{eq:lambda}). The primary exception
to the linear behaviour is that for small changes in chemoattractant
concentrations, cells respond to increases in concentration but not
decreases; we neglect this nonlinearity in our analysis.

\begin{figure}
\centering
  \includegraphics[width=0.5\textwidth]{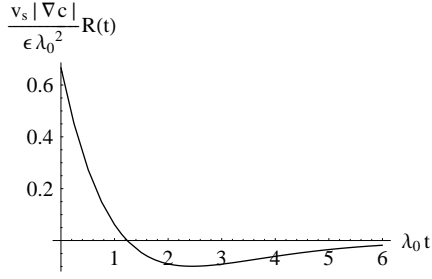}
\caption{The response function used in the results section of this
paper, originally derived by \citet{Clark.2005}.}
\label{fig:rclarkplot}       
\end{figure}

In the absence of experimental measurements of $R$, a convenient
assumption is that $R$ has a similar shape to the impulse response of
the individual motor bias reported in \citet{Block.1982} and
\citet{Segall.1986}, scaled and shifted appropriately
\citep{Schnitzer.1993}. This means that (i) $R$ should be composed of a
positive lobe followed by a negative lobe of equal area so that
\begin{equation} \label{eq:zeromean}
   \int_0^\infty dt R(t) = 0,
\end{equation}
and (ii) $R(t)$ should decay to zero for $t$ greater than about
$4\;{\rm s}$. (Note that the equivalent response function
(\ref{eq:rErban}) of the \citet{Erban.2005} model always satisfies (i)
and also satisfies (ii) provided that $t_{\rm a} < 4\,{\rm s}$.)
Additionally, we require $|\Delta(t)| \ll 1$ in order for our linear
analysis to be valid. For $R$ satisfying (\ref{eq:zeromean}),
$|\Delta(t)|$ is largest when the cell swims straight up or down the
chemoattractant gradient, with the maximum possible value of
$|\Delta(t)|$ given by
\begin{equation}
   |\Delta(t)|_{\rm max} = v_{\rm s} |\nabla c|
   \int_0^\infty dt \,t R(t).
\end{equation}
We choose $R$ to be
\begin{equation}
\label{eq:clarkR}
   R(t) = \frac{2 \epsilon \lambda_0^2}{3 v_{\rm s} |\nabla c|}
   \,\rm{e}^{-\lambda_0 t}[1-\lambda_0 t/2-(\lambda_0 t/2)^2],
\end{equation}
for which $|\Delta(t)|_{\rm max} = \epsilon$, with $\epsilon \ll 1$ in
order for our linear analysis to be valid. This $R$, plotted in figure
\ref{fig:rclarkplot}, is a theoretically motivated response function
derived in \citep{Clark.2005} that has the above-mentioned properties
(i) and (ii), and it matches the experimentally measured motor bias
reasonably well. It may seem odd that $|\nabla c|$ appears on the right
hand side of (\ref{eq:clarkR}), since $R$ is a property of the cell
rather than its environment. However the factor of $1/|\nabla c|$ in
$R$ simply reflects the fact that we have used $\epsilon$ to
parameterise the strength of the combined effect of the cell's response
$R$ and the chemoattractant gradient $\nabla c$ on $\Delta(t)$.

\subsection{Dependence of drift velocity on parameters}

\begin{figure}
\centering
  \includegraphics[width=0.6\textwidth]{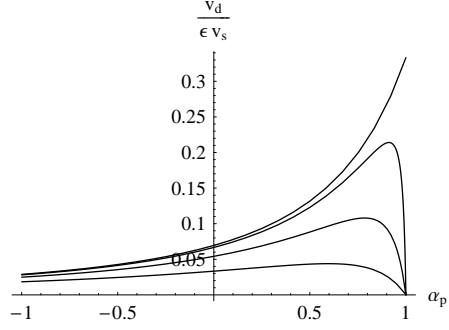}
\caption{Drift velocity $\vd$ as a function of persistence $\ap$, for
rotational diffusivities $\DR = 0,0.01,0.062,0.2 \;{\rm radians}^2{\rm
s}^{-1}$ (top to bottom), with $\lambda_0 = 1 {\rm s}^{-1}$. $\DR
=0.062\;{\rm radians}^2{\rm s}^{-1}$ is an estimate for an \emph{E.
coli} cell swimming in water at room temperature \citep{Berg.1983}.}
\label{fig:vdplot}       
\end{figure}

Substitution of (\ref{eq:clarkR}) into (\ref{eq:vdsimpleintegral})
yields
\begin{equation} \label{eq:vdclark}
   \vd =
   \frac{
      \epsilon v_{\rm s} \lambda_0^3 [\lambda_0(5-2\ap)+4\DR](1-\ap)
   }{
      9[2\DR+\lambda_0(1-\ap)][2\DR+\lambda_0(2-\ap)]^3
   }.
\end{equation}
Figure \ref{fig:vdplot} shows $\vd$ as a function of $\ap$ for a
variety of values of $\DR$.

\begin{figure}
\centering
  \includegraphics[width=0.6\textwidth]{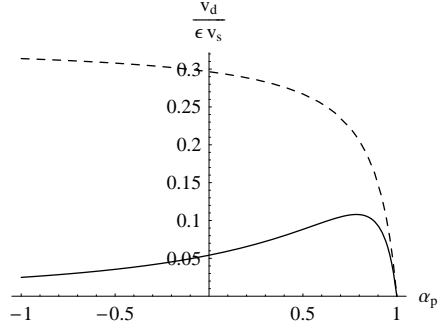}
\caption{Drift velocity $\vd$ as a function of persistence $\ap$, for
rotational diffusivity $\DR = 0.062 \;{\rm radians}^2{\rm s}^{-1}$ with
$\lambda_0 = 1 {\rm s}^{-1}$. The dashed curve is for a cell that
responds to the instantaneous chemoattractant gradient [equation
(\ref{eq:vdschnitz})] and the solid curve is for a cell that performs
temporal comparisons of chemoattractant concentration
[equation(\ref{eq:vdclark})].}
\label{fig:memvsnomem}       
\end{figure}

A striking feature of figure \ref{fig:vdplot} is that the maximum of
$\vd$ occurs at a positive value of $\ap$, so positive persistence
increases $\vd$ (up to a point). This is markedly different to the case
of a cell responding to the instantaneous chemoattractant concentration
gradient (the case analysed by \citet{Schnitzer.1993}), for which $\vd$
is highest for $\ap = -1$. Figure \ref{fig:memvsnomem} compares the two
cases. A qualitative explanation of the enhancement of $\vd$ by
persistence for a cell performing temporal comparisons is as follows.
Inspection of (\ref{eq:Delta}) reveals that the largest modulations in
$\lambda$ occur when the cell swims straight up or down the
concentration gradient for $4\;{\rm s}$ or more. For larger $\ap$, the
cell is more likely to swim approximately straight up or down the
gradient for $4\;{\rm s}$ or more (since it undergoes a smaller change
of direction when it tumbles), hence $\lambda$ is on average modulated
by a larger amount, and on average the chemotactic response is
stronger. Alternatively, one can explain the effect in terms of the
relevance of past concentration information. If $\ap$ is close to 1,
then information about the concentration over the last $4\;{\rm s}$, on
the basis of which the cell biases its tumble rate, is more relevant
than if $\ap = 0$.

Figure \ref{fig:vdplot} also shows that for $\DR>0$, $\vd \rightarrow
0$ as $\ap\rightarrow 1$. This makes sense intuitively since for $\ap =
1$ the cell effectively does not tumble at all and hence is unable to
bias its random walk. However, for $\DR = 0$ it appears that $\vd$ is
largest when $\ap = 1$, whereas sensibly one expects that $\vd=0$ when
$\ap=1$. This apparent paradox can be resolved if one considers the
time-scales over which our results are valid. In deriving
(\ref{eq:wplus}), we assumed that the probability distribution
governing $E[w_j^+]$ is steady state. This is true only if the cell has
been swimming in the chemoattractant gradient for a time $t_{\rm swim}
\gg \tau_c$, where the direction correlation time $\tau_c$ is given by
(\ref{eq:tau_c2}). For $t_{\rm swim} \gg \tau_c$, tumbles and
rotational Brownian motion have randomised the cell's swimming
direction so that the current swimming direction is independent of any
initial conditions. Clearly $\tau_c \rightarrow \infty$ as $\ap
\rightarrow 1$ for $\DR = 0$, so $\ap = 1$, $\DR = 0$ is singularity at
which our results are not valid.

The observed persistence in wild-type \emph{E. coli} is $\ap \approx
0.33$ \citep{Berg.1983}, and (\ref{eq:vdclark}) predicts that for a
rotational diffusivity of $\DR = 0.062\;{\rm radians}^2{\rm s}^{-1}$,
$\vd$ is $38\%$ larger for $\ap = 0.33$ than for $\ap =0$. Thus,
wild-type \emph{E. coli} cells enhance their drift velocity
considerably by having non-zero persistence. However, the drift
velocity is maximised by a persistence of $\ap \approx 0.78$, which is
considerably larger than the observed persistence of $\ap \approx
0.33$. This discrepancy seems surprising, since one might expect
evolution to have optimised $\ap$ to maximise $\vd$. One possible
explanation for the discrepancy is a compromise between transient and
steady-state behaviours of the cell. \citet{Clark.2005} suggested that
in order for a bacterium to be successful it must both both (a) move
toward nutrients in the short run (maximise $\vd$) and (b) in the long
run spend most of its time near the regions of highest nutrient density
if confined to a bounded domain. \citet{Clark.2005} argued that the
double lobed form of $R$ can be understood in terms of a compromise
between (a) and (b), with (a) favouring an $R$ with only a positive
lobe and (b) favouring an $R$ with only a negative lobe. A similar
compromise may govern $\ap$.

Steady state performance is optimised by an $\ap$ close to zero, for
the following reason. The translational self-diffusion coefficient of a
cell performing runs and tumbles is $D = \tau_c v_{\rm s}^2 /3$
\citep{Lovely.1975} and, from (\ref{eq:tau_c2}), $D$ is a monotonically
increasing function of $\ap$. Now, suppose a cell is confined to a
bounded domain with uniform $\nabla c$. Ideally, in the long term, the
cell would spend all its time right at the boundary with the highest
$c$. In fact, the cell wanders within a characteristic length $L$ of
that boundary, where $L \approx D/\vd$ is governed by a balance between
diffusion away from the boundary and drift velocity toward the
boundary. For $\DR = 0.062\;{\rm radians}^2{\rm s}^{-1}$, one can show
that $L$ is minimised by $\ap \approx 0.06$. A similar argument holds
if one considers a distribution of chemoattractant with a local
maximum, with the same conclusion that long term performance is
optimised by $\ap$ close to zero. Thus, the observed persistence of
$\ap \approx 0.33$ might reflect a compromise between transient and
steady-state performance.

A second factor that disfavours large $\ap$ is that \emph{E. coli} are
denser than water, which may cause them to eventually swim downward in
the absence of strong tumbles. For instance, \emph{Salmonella
typhimurium} is a denser-than-water bacterium that performs run and
tumble chemotaxis in a similar manner to \emph{E. coli}, and mutant
non-tumbling \emph{S. typhimurium} cells swim downward on average and
collect at the bottom of the vessel in which they are contained
\citep{Aswad.1975}. This is presumably because the hydrodynamic drag on
the flagella is larger than that on the cell body, so a torque is
exerted on a non-vertically-oriented cell \citep{Roberts.2002}. The
implication is that it would be disadvantageous for a cell to have
$\ap$ very close to $1$ since it would then swim downward on average,
regardless of the direction of the chemoattractant gradient.

Other factors potentially influencing the optimal value of $\ap$ for a
real \emph{E. coli} cell are that (i) an environment with background
fluid flows or with rapidly changing distributions of nutrients may
favour a smaller $\ap$, so that the cell may perform a `U-turn' more
quickly, and  that (ii) non-linear effects have been neglected in our
analysis and may play an important role for $\emph{E. coli}$ in its
native environment.

\section{Confirmation of results by Monte Carlo simulation}

Monte Carlo simulation was used to verify the correctness of the
analytic calculations for $\epsilon \ll 1$ and to assess the accuracy
of the analytic calculations for larger $\epsilon$. For the purpose of
simulations, we used a different response function:
\begin{equation} \label{eq:Rsin}
   R(t) =
   \left\{
      \begin{array}{ll}
         \epsilon
         \frac{\lambda_0^2}{v_{\rm s} |\nabla c|}
         \frac{\pi}{8}
         \sin \left( \pi\, \lambda_0\, t / 2 \right) \;& \textrm{if $0\leq \lambda_0 t\leq 4$} \\
         0 & \textrm{otherwise}
      \end{array}
   \right.
\end{equation}
This form of $R$ was chosen for computational convenience; only the
most recent four seconds of a cell's trajectory need to be stored in
memory. Like the $R$ defined by (\ref{eq:clarkR}), the $R$ defined by
(\ref{eq:Rsin}) possesses a positive lobe followed by a negative lobe
of equal area, and $|\Delta(t)|_{\rm max} = \epsilon$. Note that we do
not have to specify a value for $|\nabla c|$, since $R$ and $|\nabla
c|$ appear together as a product in the equation for $\Delta(t)$.

The simulation method was as follows. For each set of parameter values,
100 simulations were run, each simulating $10^5\, {\rm s}$ of swimming.
Time was discretised into steps of size $\Delta t = 0.01\, {\rm s}$. At
each time-step, the tumble probability was calculated using a discrete
approximation of the integral (\ref{eq:Delta}), and a random number
generator was used to decide whether the cell tumbled. If the cell
tumbled, then the new direction faced after the tumble was chosen from
an axisymmetric distribution about the old direction, such that the new
direction made an angle of $\arccos(\ap)$ with the old direction. If
the cell didn't tumble, then rotational Brownian motion was simulated
by giving the cell a new direction chosen from an axisymmetric
distribution about the old direction, such that the new direction made
an angle of $\arccos(1-2\DR\Delta t)\approx 2 \sqrt{\DR\Delta t}$ with
the old direction (note that this is consistent with equation
\ref{eq:dircort1t2}). For each simulation, the drift velocity was
estimated as the net $z$ displacement divided by the simulation
duration. The baseline tumble rate was set to $\lambda_0 = 1\,{\rm
s}^{-1}$ for all simulations.

\begin{figure}
\centering
  \includegraphics[width=0.7\textwidth]{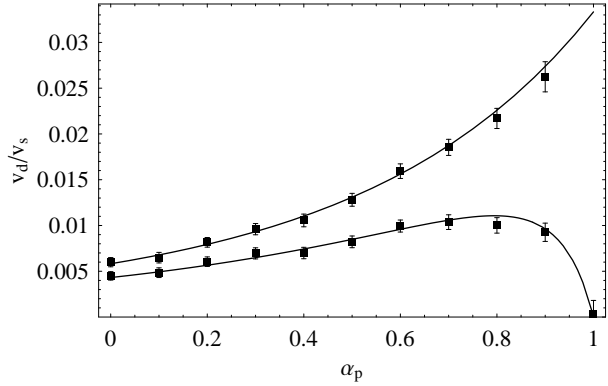}
\caption{Drift velocity $\vd$ as a function of persistence $\ap$ for
$\DR = 0$ (top) and $\DR = 0.062\,{\rm radians}^2{\rm s}^{-1}$
(bottom), with $\epsilon = 0.1$ in both cases. Solid lines show
analytic results, squares show simulation data, and error bars show
$95\%$ confidence intervals.}
\label{fig:simplots}       
\end{figure}

\begin{figure}
\centering
  \includegraphics[width=0.7\textwidth]{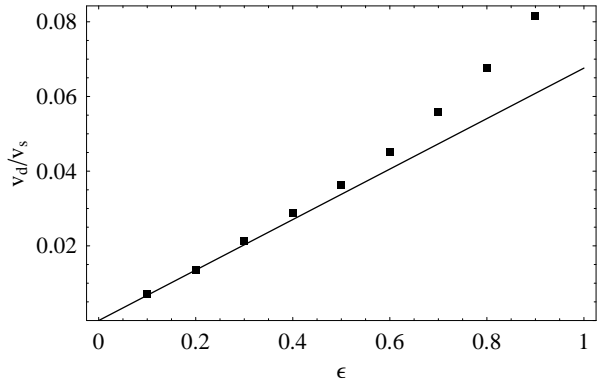}
\caption{Drift velocity $\vd$ as a function of $\epsilon$ with $\DR =
0.062\,{\rm radians}^2{\rm s}^{-1}$ and $\ap = 0.33$. The solid line
shows the analytic prediction and squares show simulation data. The
$95\%$ confidence intervals are smaller than the squares. Note that the
analytic predictions are within $10\%$ of the simulation data for
$\epsilon \leq 0.6$.}
\label{fig:epsplot}       
\end{figure}

For $\epsilon = 0.1$, there is excellent agreement between the
simulation data and analytic calculations, as shown in figure
\ref{fig:simplots}. For larger $\epsilon$, the accuracy of the analytic
calculations is diminished. Figure \ref{fig:epsplot} shows that the
results of the analytic calculations are within $10\%$ of the
simulation results for $\epsilon$ up to approximately $0.6$.

\section{Discussion and Conclusion}

In this paper we presented an analytic method for calculating the drift
velocity of a cell performing run and tumble chemotaxis, taking account
of (i) temporal comparisons, (ii) persistence, and (iii) rotational
Brownian motion, and we verified the results with Monte Carlo
simulations. The calculations are novel in that they are the first
calculation of chemotactic drift velocity simultaneously to include all
features (i), (ii) and (iii) of the motion. Using estimates for the
chemotactic response function and the rotational diffusivity, our key
finding is that persistence can markedly increase the drift velocity.
For instance, a persistence of $\ap = 0.33$ (characteristic of
wild-type \emph{E. coli} \citep{Berg.1983}) can increase the drift
velocity by $\approx 40\%$ relative to $\ap = 0$, while a persistence
of $\ap = 0.78$ can double the drift velocity relative to $\ap = 0$.

There are a number of possible reasons for discrepancy between the
value of $\ap$ that maximises $\vd$ and the value of $\ap$ observed in
wild-type \emph{E. coli}. First, our model looks only at behaviour in
an unbounded domain, or equivalently at transient behaviour in a
bounded domain, whereas one expects that in order to be successful a
bacterium must optimise both its transient and its steady state
behaviour. Indeed, it has been proposed that the double-lobed form of
$R$ reflects a compromise between transient and steady state behaviour
\citep{Clark.2005}. The same might be true of $\ap$, with a smaller
value of $\ap$ resulting in a more favourable steady-state behaviour.
Second, our analysis neglects the fact that \emph{E. coli} cells are
denser than the fluid in which they swim and geometrically asymmetric
(front-to-back), so that a cell with $\ap \approx 1$ would swim
downward on average. Third, our analysis does not include rapidly
varying chemoattractant distributions or background fluid flows.
Finally, our analysis is linear and therefore limited to weak
chemoattractant gradients; it is not inconceivable that the dependence
of drift velocity on persistence might be different in stronger
chemoattractant gradients.

\begin{acknowledgements}
I would like to thank Professor Tim Pedley for his guidance and
encouragement, Dr Kalvis Jansons for his advice on notation, and
Professor Howard Berg for fruitful conversations.
\end{acknowledgements}

\bibliographystyle{spbasic}
\bibliography{everything}

\appendix

\section{Solving the Fokker-Planck equation}
\label{app:fpe}

The Fokker-Planck equation is
\begin{equation}
\label{eq:appfpe}
   \frac{\partial f}{\partial t} = \DR \nabla_{\bs{e}}^2 f,
\end{equation}
where $\nabla_{\bs{e}}^2$ is the Laplacian in direction space and
$\bs{e}$ is the swimming direction. We can describe $\bs{e}$ by
spherical polar coordinates, so that $\bs{e}\cdot\bs{e}_z = \cos\theta$
and $ f(\theta,\phi) \sin\theta \,d\theta\, d\phi $ is the probability
that $\bs{e}$ lies in the solid angle $\sin\theta \, d\theta \, d\phi$.
Assume axisymmetry, so that $f$ is a function of $\theta$ only. Writing
out the Laplacian in full, we then have
\begin{equation}
\label{eq:appfpe2}
   \frac{\partial f}{\partial t}= \DR \left(
   \frac{\partial^2 f}{\partial\theta^2}
   +\cot\theta \,\frac{\partial f}{\partial\theta} \right).
\end{equation}
Without loss of generality, we choose our initial condition to be
$\bs{e}(0) = \bs{e}_z$. Since we are only interested in the direction
correlation function, we need only calculate $E[ \cos\theta(t) ]$
rather than solve for $f$. Consider
\begin{equation}
\label{eq:dcosthetadt}
   d E[ \cos\theta]/dt = 2\pi \int_0^\pi \sin\theta\,
   \cos\theta \,\frac{\partial f}{\partial t}\, d\theta.
\end{equation}
Substituting (\ref{eq:appfpe2}) into (\ref{eq:dcosthetadt}) and
integrating by parts, one finds
\begin{equation}
   d E[ \cos\theta]/dt = - 2 \DR \langle \cos\theta \rangle,
\end{equation}
so
\begin{equation}
   E[ \cos\theta(t) ] = {\rm e}^{-2\DR t}.
\end{equation}
Generalising from this result it is straightforward to show that
\begin{equation}
   E[ \bs{e}(t_1) \cdot \bs{e}(t_2) ] = {\rm
   e}^{-2\DR|t_2-t_1|}.
\end{equation}


\end{document}